\documentclass[a4paper,11pt]{article}
\pdfoutput=1
\usepackage{jheppub}
\usepackage[T1]{fontenc}
\usepackage{amssymb}
\usepackage{amsmath}
\usepackage{graphicx}
\usepackage{tabularx}
\usepackage{hyperref}

\title{\boldmath The Gravitational-Wave Physics}

\author[1,3]{Rong-Gen Cai}
\author[2,4]{Zhoujian Cao}
\author[1,5]{Zong-Kuan Guo}
\author[1,3]{Shao-Jiang Wang}
\author[1,3]{Tao Yang}

\affiliation[1]{CAS Key Laboratory of Theoretical Physics, Institute of Theoretical Physics, Chinese Academy of Sciences, No.55 Zhong Guan Cun East Road, Beijing 100190, China}
\affiliation[2]{Department of Astronomy, Beijing Normal University, No. 19, XinJieKouWai Street, Beijing 100875, China}
\affiliation[3]{School of Physical Sciences, University of Chinese Academy of Sciences, No.19A Yuquan Road, Beijing 100049, China}
\affiliation[4]{Institute of Applied Mathematics, Academy of Mathematics and Systems Science, Chinese Academy of Sciences, No.55 Zhong Guan Cun East Road, Beijing 100190, China}
\affiliation[5]{School of Astronomy and Space Science, University of Chinese Academy of Sciences, No.19A Yuquan Road, Beijing 100049, China}

\emailAdd{cairg@itp.ac.cn}
\emailAdd{zjcao@amt.ac.cn}
\emailAdd{guozk@itp.ac.cn}
\emailAdd{schwang@itp.ac.cn}
\emailAdd{yangtao@itp.ac.cn}

\abstract{The direct detection of gravitational wave by Laser Interferometer Gravitational-Wave Observatory indicates the coming of the era of gravitational-wave astronomy and gravitational-wave cosmology. It is expected that more and more gravitational-wave events will be detected by currently existing and planned gravitational-wave detectors. The gravitational waves open a new window to explore the Universe and various mysteries will be disclosed through the gravitational-wave detection, combined with other cosmological probes. The gravitational-wave physics is not only related to gravitation theory, but also is closely tied to fundamental physics, cosmology and astrophysics. In this review article, three kinds of sources of gravitational waves and relevant physics will be discussed, namely gravitational waves produced during the inflation and preheating phases of the Universe, the gravitational waves produced during the first-order phase transition as the Universe cools down and the gravitational waves from the three phases: inspiral, merger and ringdown of a compact binary system, respectively. We will also discuss the gravitational waves as a standard siren to explore the evolution of the Universe.
\begin{flushleft}
\end{flushleft}
}

\begin{document}
\maketitle
\flushbottom

\section{Introduction}
\label{sec:introduction}

On 11 February 2016, the LIGO Scientific Collaboration and the Virgo Collaboration~\cite{Abbott:2016blz} announced that on 14 September 2015 at 09:50:45 UTC the two detectors of the Laser Interferometer Gravitational-Wave Observatory (LIGO) simultaneously observed a transient gravitational-wave (GW) signal. The GW event is named GW150914. The frequency of the signal increases from 35 to 250 Hz with a peak GW strain $1.0\times 10^{-21}$ at the $150$ Hz.  The GW signal is consistent with the one predicated by general relativity for the inspiral and merger of a pair of black hole and the ringdown of the resulting single black hole. The source is located at a luminosity distance of $410^{+160}_{-180}$ Mpc corresponding to a redshift $z=0.09^{+0.03}_{-0.04}$. The initial black hole masses are $36^{+5}_{-4}M_{\odot}$ and $29^{+4}_{-4}M_{\odot}$ and the resulting black hole mass is $62^{+4}_{-4}M_{\odot}$. The mass difference $3.0^{+0.5}_{-0.5}M_{\odot}$ is radiated in the form of GWs. This is the first direct detection of GW and the first observation of a binary black hole merger. On 15 June 2016, the second GW event, GW151226, was announced by the same team~\cite{Abbott:2016nmj}.  This time, observed signal lasts approximate 1 s, the frequency increases from 35 to 450 Hz within 55 cycles, and the peak gravitational strain reaches $3.4^{+0.7}_{-0.9} \times 10^{-22}$ at $450$ Hz. The source is also the merger of two black holes with masses $14.2^{+8.3}_{-3.7}M_{\cdot}$ and $7.5^{+2.3}_{-2.3}M_{\odot}$, respectively, the final black hole has mass $20.8^{+6.1}_{-1.7}M_{\odot}$. This event happened with a luminosity distance $440^{+180}_{-190}$ Mpc corresponding to a redshift $0.09^{+0.03}_{-0.04}$.

The GW was predicted by Albert Einstein in 1916~\cite{Einstein:1916cc,Einstein:1918btx}, 1 year later after he finally formulated his theory on gravitation, genera relativity. But the physical reality of the GW solution of the Einstein field equations was not showed until the Chapel Hill conference in 1957~\cite{Saulson:2010zz}. In \cite{Bondi:1957dt,Bondi:1958aj}, it has been shown that GW carries energy and when passing through the spacetime in a form of a sandwich, it affects test particles. More than one century has passed since the Einstein's proposal of general relativity, although it passed various precise tests, some alternatives still survive, for example, scalar-tensor gravity theory, $f(R)$ gravity, modified gravity with higher curvature terms, etc. On the other hand, based on general relativity, one has now a standard model (SM) of cosmology, $\Lambda$CDM model, which is quite well consistent with various astronomical  observations made so far. Now we understand well that GW exists not only in general relativity, but also in other  relativistic covariance gravity theories. Due to the limited space, this review is confined for the GWs in general relativity.

The sources of GWs could be classified into two categories roughly. One is called cosmological origin, the other is relativistic astrophysical origin. In the cosmological case, GWs can be produced in the early stages of the Universe, for example, during the inflation and reheating epochs. Such GWs are called primordial GWs, and they will leave unique imprint on the cosmic microwave background (CMB), the so-called $B$-mode. On the other hand, during the evolution of the Universe, it is expected that various phase transitions (PTs) had happened as the temperature of the Universe decreases, for example, the symmetric breaking of the grand unification theory, electroweak (EW) PT, quantum chromodynamics (QCD) PT, etc. During those PTs, GWs are also expected to be produced with different features.  Also the interactions of topological defects such as cosmic string, domain wall, etc, produced during PTs, will create GWs. Therefore, the detection of GWs due to those cosmological origins can reveal physics associated with the evolution of the Universe. In the astrophysics side, GWs can be produced in various processes, for example, rotation of non-symmetric neutron star, explosion of supernovae, inspiral, merger and ringdown of some compact binaries including white dwarf, neutron star and/or black hole. In particular, the compact binary systems are main sources for GW detection such as LIGO, Virgo, Kagra, Einstein Telescope, etc., ground-based GW experiments, and Laser Interferometer Space Antenna (LISA), Deci-Hertz Interferometer Gravitational wave Observatory (DECIGO), Big Bang Observer (BBO), Taiji, Tianqin, etc., space-based GW experiments.

Therefore, the GW physics is closely related to fundamental physics, cosmology and astrophysics. The direct detection of GWs in~\cite{Abbott:2016blz,Abbott:2016nmj} indicates the coming of the era of GW astronomy and GW cosmology. The detection of GWs opens a new window to explore the Universe. Combining the electromagnetic (EM) radiation, neutrino, cosmic ray and GWs, it could be expected that one is able to reveal various currently existing mysteries concerning the early evolution of the Universe, property of dark matter, the nature of dark energy, etc. In this brief review, we are going to summarize some important aspects relevant to the GW physics.

The outline of this review is as follows. Section~\ref{sec:earlyuniverse} is going to introduce the GWs produced in the primordial Universe and its detection through CMB. In particular, we emphasize the properties of GWs created during inflation and preheating processes and current situation of detection of the primordial GWs. In section~\ref{sec:PT} we will discuss the GWs produced during cosmic PT. There we will introduce the bubble nucleation, bubble expansion and bubble percolation. During the strong first-order PT, bubble collision, turbulent magnetohydrodynamics (MHD) and sound wave are all the sources to produce GWs. Section~\ref{sec:binary system} will mainly be devoted to discussions on the GWs from the dynamics of compact binary systems. There we will introduce three main methods to solve the binary system: the post-Newtonian (PN) approximation, numerical relativity and the black hole perturbation, corresponding three phases: inspiral, merger and ringdown of two black holes system, respectively. In that section, we will also discuss the possibility of GWs produced by binary systems as a cosmological probe. With this new probe, it is expected to have a strong constraint on cosmological parameters, combining with other cosmological probes.

\section{Gravitational Waves From Primordial Universe}
\label{sec:earlyuniverse}

\begin{figure}
  \centering
  \includegraphics[width=0.9\textwidth]{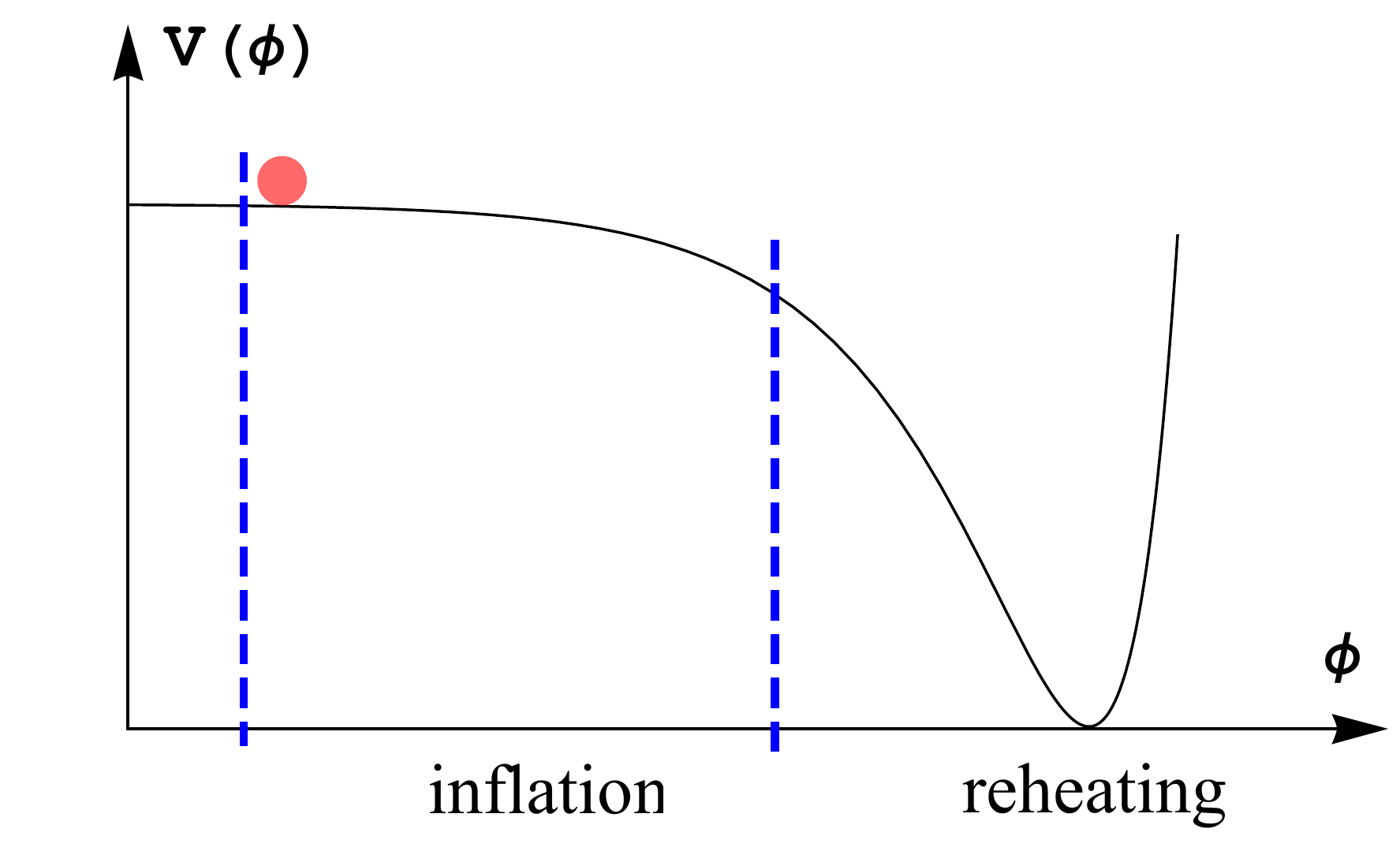}
  \caption{The schematic illustration of the inflaton potential. In the slow-roll inflationary scenario, an accelerated expansion of the Universe
  occurs when the inflaton rolls slowly along its potential. After inflation ends, the inflaton oscillates around the minimum of its potential, whose  energy is converted into radiations.}
\label{fig:potential}
\end{figure}

In order to solve some problems in big-bang cosmology such as the horizon and flatness problems, inflationary scenario was introduced~\cite{Guth:1980zm,Albrecht:1982wi,Linde:1981mu}, in which a period of accelerated expansion of the Universe happened at early times.
Inflation not only predicts the primordial scalar perturbations, which provide a natural way to generating the anisotropies of the CMB radiation and the initial tiny seeds of the large-scale structure observed today in the Universe, but also generates a stochastic background of the primordial GWs. Although such a stochastic background of the primordial GWs has not been observed yet, its detection would open a new window to understanding the physics of the early Universe and thus the origin and evolution of the Universe. In this section, we shall firstly review the properties of the primordial GWs produced during inflation and preheating (see Fig.~\ref{fig:potential}), and then discuss observational implications.

GWs are described by a transverse-traceless gauge-invariant tensor perturbation, $h_{ij}$, in a Friedman-Robertson-Walker (FRW) metric,
\begin{eqnarray}
\mathrm{d}s^2=a^2(\tau)\left[-\mathrm{d}\tau^2+(\delta_{ij}+h_{ij})\mathrm{d}x^i\mathrm{d}x^j\right],
\end{eqnarray}
where $\tau$ is the conformal time and $a$ is the scale factor, and $h_{ij}$ satisfies $\partial^i h_{ij}=0$ and $\delta^{ij}h_{ij}=0$. To first order in $h_{ij}$, the perturbed Einstein equation reads
\begin{eqnarray}
\label{eq:pee}
h_{ij}^{\prime\prime}+2{\cal{H}}h_{ij}^{\prime}-\triangledown^2h_{ij}=\frac{2}{M_{\rm pl}^2}\Pi_{ij}^{\rm TT},
\end{eqnarray}
where the prime denotes the derivative with respect to $\tau$, ${\cal{H}} \equiv a'/a$ is the Hubble parameter in $\tau$, $M_{\rm pl} \equiv (8\pi G)^{-1/2}$ is the reduced Planck mass, and the source term; $\Pi_{ij}^{\rm TT}$ is the transverse-traceless projection of the anisotropic stress tensor $T_{ij}$. Since $h_{ij}$ is symmetric, transverse and trace-free, tensor modes are left with two physical degrees of freedom, which are
expanded in the Fourier space as
\begin{eqnarray}
h_{ij}(\tau,{\mathbf x}) = \int \frac{\mathrm{d}^3{\mathbf k}}{(2\pi)^{3/2}} e^{i{\mathbf k}\cdot{\mathbf x}} \left[h^{+}_{\mathbf k}(\tau)e_{ij}^{+}(\mathbf x)+h^{\times}_{\mathbf k}(\tau)e_{ij}^{\times}(\mathbf x)\right],
\end{eqnarray}
where $e_{ij}^{+,\times}$ are the polarization tensors with two polarization states ($+,\times$) of GWs. We have the power spectrum of GWs as
\begin{eqnarray}
P_T(\tau,k)=\frac{k^3}{2\pi^2}(|h^+_{\mathbf k}|^2+|h^\times_{\mathbf k}|^2),
\end{eqnarray}
and the energy spectrum of GWs as $\Omega_{\rm gw}(\tau,k) \equiv \mathrm{d}\rho_{\rm gw}/\mathrm{d}\ln k/\rho_c$, where
\begin{eqnarray}
\rho_{\rm gw}(\tau,k)=\frac{1}{4a^2}M_{\rm pl}^2 \langle h^{\prime}_{ij}h^{\prime ij}\rangle
\end{eqnarray}
is the energy density and $\rho_c=3H^2M^2_{\rm pl}$ is the critical density of the Universe.

\subsection{During Inflation}
\label{subsec:inflation}

In the standard single-field slow-roll inflationary scenario, at first order in perturbation theory Eq.~\eqref{eq:pee} reduces to a free wave equation. In this case, the wave equation can analytically be solved in the slow-roll approximation. The Bunch-Davies vacuum condition in the asymptotic past is imposed because the modes lie well inside the Hubble radius. Of course, if a general vacuum condition is imposed, there are additional features in the power spectrum. During inflation, quantum fluctuations are amplified and stretched, and then nearly frozen on super-Hubble scales. The single-field slow-roll inflation predicts a slightly red-tilted spectrum:
\begin{eqnarray}
P_{T}(k)=\frac{8}{M^2_{\rm pl}}\left(\frac{H}{2\pi}\right)^2,
\end{eqnarray}
and a consistency relation $n_{T}=-r/8$ between the tensor spectral index $n_{T}$ and the tensor-to-scalar ratio $r$. Here, $n_T$ and $r$ are evaluated at the epoch when a given perturbation mode leaves the Hubble horizon, {\it i.e.} $k=aH$. We usually introduce the tensor-to-scalar ratio $r\equiv A_{T}/A_{S}$ at a pivot scale $k_{\ast}$, {\it i.e.} the amplitude of tensor perturbations $A_{T}$ with respect to that of scalar perturbations $A_{S}$. The red-tilted spectrum means that the amplitude of tensor perturbations becomes small on small scales due to the fact that large-scale modes are earlier stretched across the Hubble horizon than small-scale modes as the energy density slowly decreases during inflation. From the wave equation we see that the evolution of $h_{ij}$ depends explicitly on $a(\tau)$ and implicitly on inflation potential through FRW equations. Hence the power spectrum of GWs encodes useful information on the evolution of the scale factor. Moreover, the tensor-to-scalar ratio is related to the energy scale of inflation by $V=3\pi^2M_{\rm pl}^4A_sr/2=(1.88\times10^{16} {\rm GeV})^4 r/0.10$. Here we have adopted the estimated value of the amplitude of scalar perturbations from the Planck 2015 data~\cite{Ade:2015lrj}. Different models predict different values of the tensor-to-scalar ratio. For example, the simplest chaotic inflation with a quartic potential~\cite{Linde:1981mu} predicts a large value of $r\approx 0.26$ while the $R^2$ inflation~\cite{Starobinsky:1980te} predicts a small value of $r\approx 0.0033$ to lowest order in slow-roll parameters if  the number of e-folds $N_{\ast}=60$ is assumed. With the help of the scalar spectral index, the estimated value of $r$ is robust to discriminate slow-roll inflationary models. In summary, the measurement of the power spectrum of GWs helps us to
\begin{itemize}
\item test the vacuum initial condition,
\item detect the evolution of the scale factor,
\item determine the energy scale of inflation,
\item discriminate inflationary models.
\end{itemize}

If the primordial GWs are detected, the next important question to answer is what is the shape of the power spectrum of GWs and whether there are additional features in the power spectrum. In the slow-roll inflationary scenario, the shape of the tensor power spectrum is characterized by the tensor spectral index $n_T$ since the running of the spectral index is negligible to the  lowest order in slow-roll parameters.
More general shapes beyond slow-roll may be reconstructed by using a binning method of a cubic spline interpolation in a logarithmic wavenumber space~\cite{Guo:2011re,Guo:2011hy,Hu:2014aua}. Checking the consistency relation between $n_T$ and $r$ provides a powerful test of the single-field slow-roll inflationary scenario. The violation of the consistency relation could in principle come from the following two aspects. The first  is that the second and third terms on the left hand side of Eq.~\eqref{eq:pee} are modified in general inflationary models, such as the k-inflation~\cite{ArmendarizPicon:1999rj}, Gauss-Bonnet inflation~\cite{Guo:2009uk,Guo:2010jr,Jiang:2013gza} and generalized G-inflation~\cite{Kobayashi:2011nu}. In the model of k-inflation, the consistency relation becomes $n_{T}=-r/8c_S$, where $c_S$ is the sound speed of scalar perturbations. In the Gauss-Bonnet inflationary model, the consistency relation is broken due to $n_{T}=-r/8-\delta_1$, where $\delta_1$ is determined by the Gauss-Bonnet coupling term. In the model of generalized G-inflation, the tensor spectral index not only depends on the evolution of the scale factor but also the higher-order derivative terms, which admits a blue-tilted power spectrum of GWs. The second is that there exists a non-negligible source term on the right hand side of Eq.~\eqref{eq:pee} during inflation. In this case, the wave equation is solved by the Green's function method. Possible sources of generating GWs include first-order scalar perturbations~\cite{Matarrese:1997ay}, perturbations of the extra field such as the curvaton~\cite{Bartolo:2007vp} and spectator field~\cite{Biagetti:2013kwa,Biagetti:2014asa}, and particle production during inflation~\cite{Cook:2011hg}.

\subsection{During Preheating}
\label{subsec:preheating}

In the inflationary scenario, at the end of inflation, the inflaton field begins to oscillate around the minimum of its potential. Such coherent oscillations produce elementary particles and eventually reheat the Universe. This process is called reheating~\cite{Albrecht:1982mp}. The coupling between the inflaton field and other fields is necessarily tiny ensuring that reheating proceeds slowly. Preheating provides a more rapidly efficient mechanism for extracting energy from the inflation field by parametric resonance~\cite{Traschen:1990sw}. Such a process is so rapid that the produced particles are not in thermal equilibrium. The preheating leads to large and time-dependent inhomogeneities of the stress tensor that source a stochastic background of GWs~\cite{Khlebnikov:1997di}. Unlike GWs produced during inflation, they are generated and remain in the Hubble horizon until now. Clearly, their wavelengths are smaller than the Hubble radius at the time of GW production. Therefore, the peak frequency of this type of stochastic GWs is typically of order more than $10^3$ Hz. Detecting such high-frequency GWs is particularly challenging. For example, for the $\phi^4$ and $\phi^2$ chaotic inflationary models, lattice simulations~\cite{Felder:2000hq} show that preheating can lead to GWs with frequencies of around $10^6 \sim 10^8$ Hz and peak power of $\Omega_{\rm gw}h^2 \approx 10^{-9} \sim 10^{-11}$ at present~\cite{Easther:2006gt}. For hybrid inflation, GWs cover a larger range of frequencies. The peak wavelength depends essentially on the coupling constant~\cite{Easther:2006vd}. See~\cite{Guzzetti:2016mkm} for the most recent review on the GWs from the inflationary era and preheating era.

\subsection{Observational Implications}
\label{subsec:observation}

\begin{figure}
  \centering
  \includegraphics[width=0.9\textwidth]{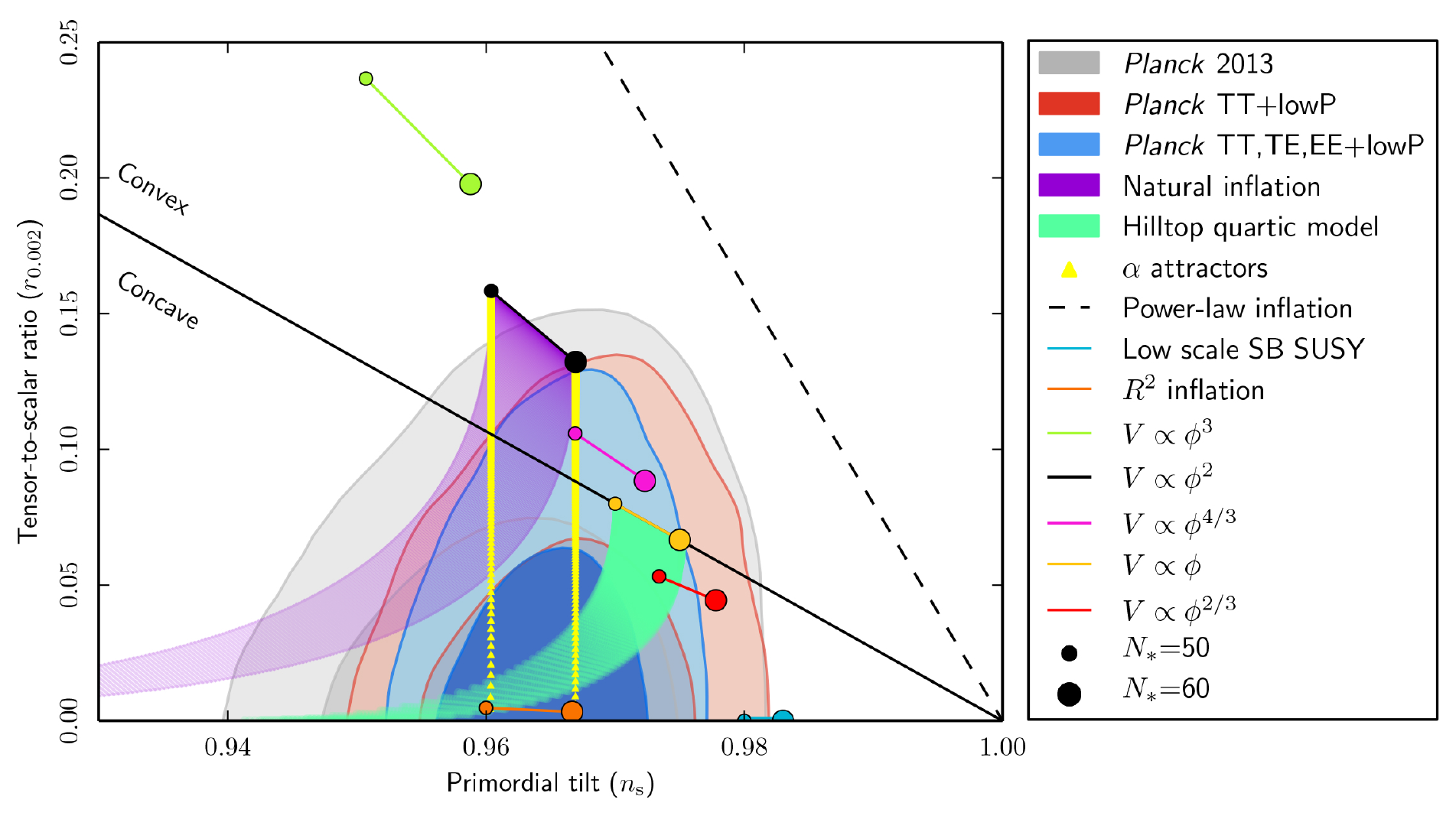}
  \caption{Marginalized joint 68\% and 95\% confidence level regions for $n_s$ and $r_{0.002}$ from the Planck 2015 data, compared
  to the theoretical predictions of some inflationary models. This figure is quoted from~\cite{Ade:2015lrj}.}
\label{fig:planck2015}
\end{figure}

Observational constraints on GWs produced during inflation mainly come from the B-mode polarization of the CMB anisotropies. Such GWs generate a quadrupolar anisotropy of momentum $m=2$ in the intensity field of photons at the epoch of recombination while scalar perturbations generate only a quadrupolar anisotropy of momentum $m=0$. Importantly, only the quadrupolar anisotropy of momentum $m=2$ causes the B-mode polarization. Therefore, the measurement of the B-mode polarization allows us to probe GWs produced during inflation. The Planck 2015 data give the 95\% confidence level upper bound for the tensor-to-scalar ratio $r_{0.002} < 0.10$ at the pivot scale $k_{\ast}=0.002$ Mpc$^{-1}$~\cite{Ade:2015lrj}, which is improved to $r_{0.002} < 0.08$ by adding the BKP cross-correlation likelihood~\cite{Ade:2015tva}. With the help of the scalar spectral index $n_S$, slow-roll inflationary models are discriminated in the $r-n_S$ plane, as shown in Fig.~\ref{fig:planck2015}.
For example, the inflationary model with a quartic potential~\cite{Linde:1981mu} is strongly disfavored by recent CMB data. Future ground-based and space-based CMB experiments will provide high precision measurements of the B-mode polarization. Actually, constraints on GWs produced during inflation from measurements of the CMB anisotropies are limited to a narrow scale range of $10^{-4} \sim 10^{-1}$ Mpc$^{-1}$, which corresponds to the change of e-folds number $\Delta N \approx 8$. This implies that it is impossible to detect the global shape of the tensor spectrum by using only CMB data. Laser interferometer gravitational-wave experiments provide the possibility of measuring the stochastic background at small scales. During inflation, quantum fluctuations of $h_{ij}$ are amplified and stretched across the Hubble horizon, and then nearly frozen on super-Hubble scales. After reentering the Hubble horizon at the epoch of radiation or matter dominance, tensor perturbations begin to oscillate with the amplitude damped by a factor $a^{-1}$. Unfortunately, for the nearly scale-invariant tensor spectrum predicted by the single-field slow-roll inflationary models, the energy spectrum in the frequency range of $10^{-10} \sim 10^3$ Hz at present becomes too weak to be directly detected by pulsar timing array (PTA) experiments and laser interferometer experiments~\cite{Boyle:2005se}. If it is a blue-tilted tensor spectrum predicted by inflationary models beyond slow-roll or caused by the source term, the direct detection might be possible in the future~\cite{Bartolo:2016ami}.

For GWs produced during preheating, since their wavelengths are smaller than the Hubble radius at the time of GWs production, the corresponding peak frequencies are typically of order more than $10^3$ Hz. It is particularly challenging to detect such high-frequency GWs by ground-based laser interferometer experiments.

\section{The Gravitational Waves From Phase Transitions}
\label{sec:PT}

It is believed that our observable Universe has experienced several PTs (PTs), during which an unknown high-degree symmetry is subsequently relaxed down to the broken EW symmetry that is well described by the SM of particle physics. If the PT from a high-temperature symmetric phase to a low-temperature symmetry-broken phase is of first order, the true vacuum bubbles would be nucleated within the false vacuum, leading to the expanding, colliding and merging bubbles that generate a stochastic background of GWs (see~\cite{Caprini:2015zlo} for recent review and~\cite{Binetruy:2012ze} for the discussions of GWs from cosmic strings and domain walls, which will not be discussed in this review). The primary motivations to study the GWs from PTs are two-folds: First, the EWPT of SM is cross-over according to the current measurements of Higgs mass. Therefore, any detections of GWs from PTs would necessarily provide us a unique probe beyond the SM, which cannot be directly probed by the particle colliders in a foreseeable future; second, the stochastic backgrounds of GWs also consist of the GWs from the inflation era and the reheating era and the other cosmological defeats from PTs like cosmic strings and domain walls. Therefore, it would help us to extract the GWs of astrophysical sources from the stochastic backgrounds of GWs. However, the topics of detections are not discussed in this review, which should merit another paper to discuss how to detect the stochastic background of GWs and how to distinguish the signals of PTs from those signals of reheating or other cosmological defeats.

\subsection{Bubble Nucleation}
\label{subsec:bubble nucleation}

In the seminal papers~\cite{Coleman:1977py,Callan:1977pt}, the first semiclassical description of vacuum decay in flat spacetime was developed for a single scalar field without derivative interactions. The vacuum decay is implemented through barrier penetration from the unstable false vacuum to the stable true vacuum, of which the field configuration is captured in the so-called bounce equation. The bounce equation describes the true vacuum bubbles nucleated during barrier penetration in the surrounding false vacuum with the probability per unit time per unit volume being of form $\Gamma=A\exp(-B/\hbar)[1+\mathcal{O}(\hbar)]$. The coefficient $B$ was worked out in~\cite{Coleman:1977py} as the on-shell Euclidean  action  of the bounce solution and the coefficient $A$ was worked out in~\cite{Callan:1977pt} to properly account for the quantum corrections \footnote{The precise form of $A$ is not important in the study of GW from PTs because $B$ dominates the evolution of decay probability though its exponential dependence.}. In the very special case where the potential barrier is larger than the difference of energy density between false and true vacuums, a thin-wall approximation was proposed in~\cite{Coleman:1977py} to evaluate the bounce action in a closed form consisting both contributions from the bubble interior and the bubble wall. The insight from thin-wall approximation provides us a physical picture of bubble nucleation that the released energy from spreading true vacuum into false vacuum goes to the acceleration of the bubble wall until a critical bubble with zero energy is formed. The size of such critical bubble was determined by the stationary point of the Euclidean action, which gives a rough cancelation between the energy from bubble interior and the energy from bubble wall.

\begin{figure}
  \centering
  \includegraphics[width=0.9\textwidth]{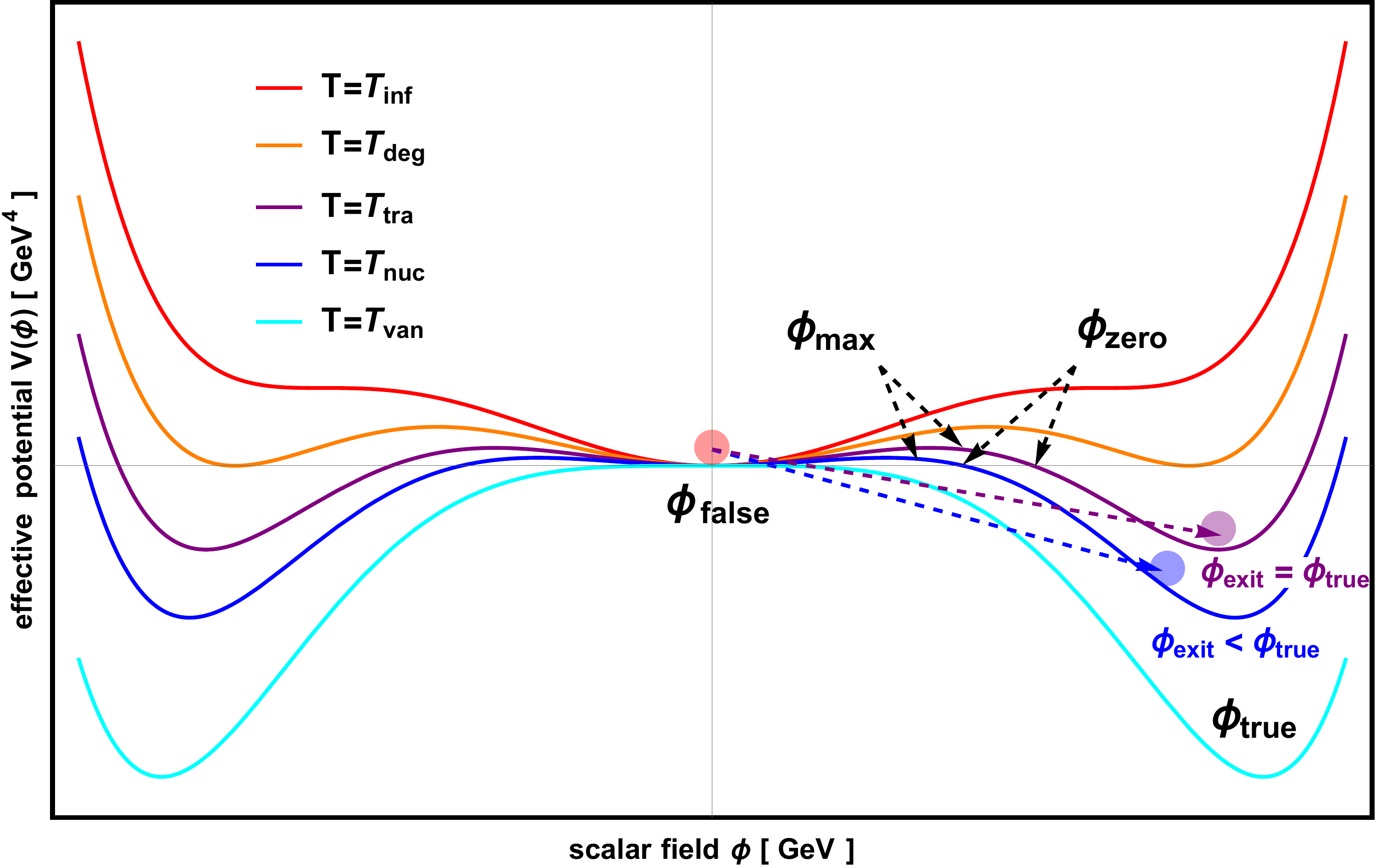}\\
  \includegraphics[width=0.9\textwidth]{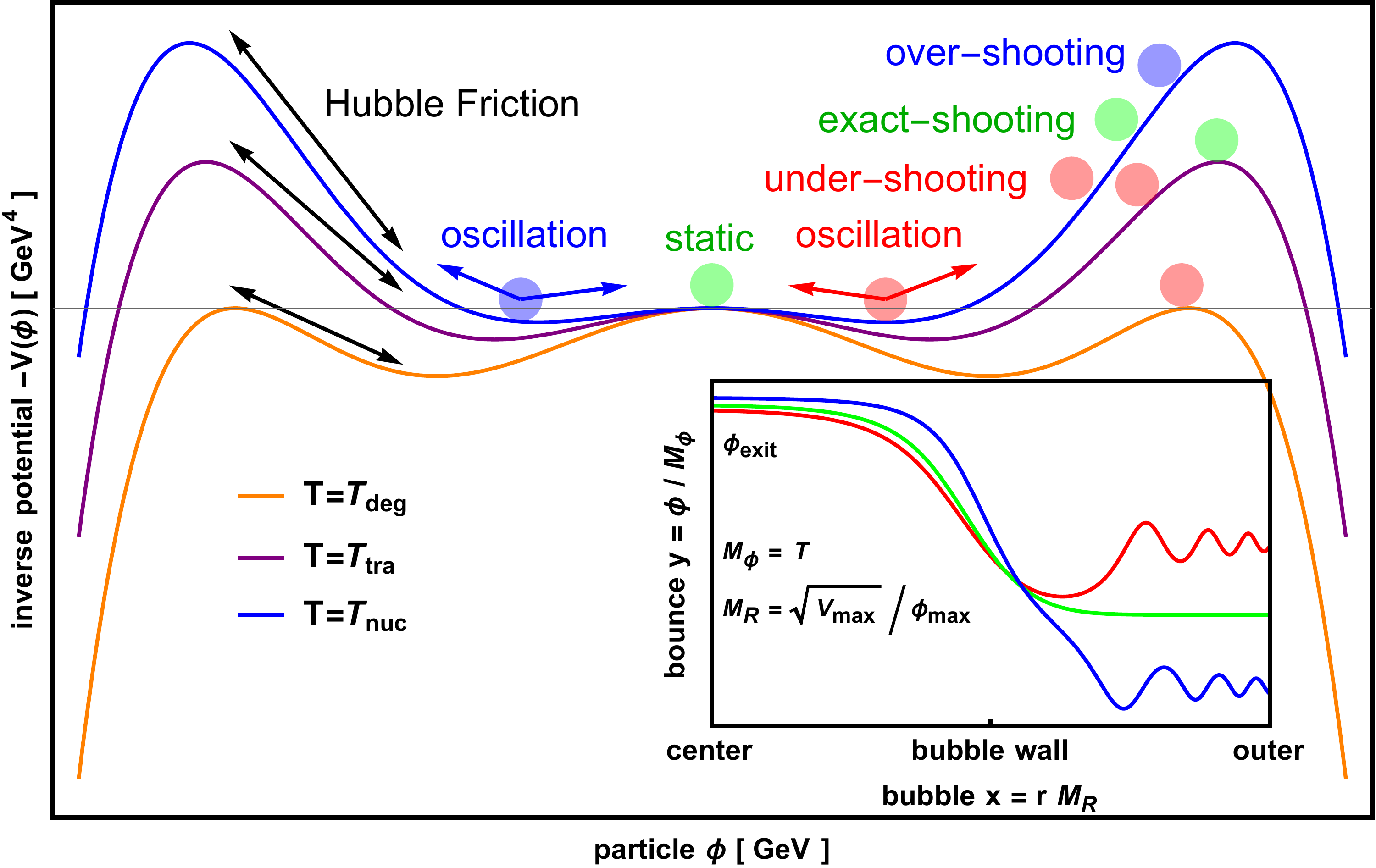}\\
  \caption{The schematic illustration of shooting algorithm for the bounce equation, which is equivalent to a particle moving in the inverse potential with Hubble friction. The exit point of field profile $\phi_\mathrm{exit}$ presents the field value at the center of nucleated bubble, which can be found by the shooting algorithm with errors and trials of under/over shooting until an exact shooting is achieved. Here, the bubble profile is rescaled by the temperature $M_\phi$ with respect to the bubble size rescaled by the mass scale $M_R$ of effective potential.}
  \label{fig:decayshooting}
\end{figure}

The seminal works~\cite{Coleman:1977py,Callan:1977pt} were later extended to the case of vacuum decay in curved spacetime in~\cite{Coleman:1980aw} and thermal decay in flat spacetime in~\cite{Linde:1981zj}. Although the presence of gravity does not change the general picture, it does lower the probability of vacuum decay by nucleating larger bubbles. In the extreme case of almost degenerated vacuums, gravity can even stabilize the false vacuum by preventing us from making a zero-energy bubble in the false vacuum, which in the absence of gravity can be simply done by expanding the bubble interior to balance the surface tension of bubble wall. When the temperature turns on, instead of an $O(4)$-symmetric bubble in Euclidean spacetime, an $O(3)$-symmetric bubble solution that is periodic in time direction with period of inverse temperature $T^{-1}$ is expected, and the size of such critical bubble is determined by the maximum point of total energy. The tunneling rate was estimated in~\cite{Linde:1980tt,Linde:1981zj} as
\begin{align}
\Gamma(T)\simeq T^4\left(\frac{S_3[\phi_B(r),T]}{2\pi T}\right)^\frac32 \exp\left(-\frac{S_3[\phi_B(r),T]}{T}\right),
\end{align}
where the Euclidean action $S_3[\phi(r),T]$,
\begin{align}
S_3[\phi(r)]=4\pi\int_0^\infty\mathrm{d}r r^2\left[\frac12\left(\frac{\mathrm{d}\phi}{\mathrm{d}r}\right)^2+V(\phi,T)\right],
\end{align}
is estimated at the bounce profile $\phi_B(r)$ of equation-of-motion,
\begin{align}
\frac{\mathrm{d}^2\phi}{\mathrm{d}r^2}+\frac{2}{r}\frac{\mathrm{d}\phi}{\mathrm{d}r}=\frac{\partial V}{\partial\phi}.
\end{align}
The effective potential usually consists of tree-level potential, zero-temperature corrections and finite-temperature corrections including the daisy resummation. Recently, a new thermal resummation procedure was proposed in~\cite{Curtin:2016urg}, which makes it possible to match theories to an EFT at finite temperature. Future works can be carried out along this direction. The naive strategy of solving above bounce equation is the shooting algorithm, which is illustrated in  Fig.~\ref{fig:decayshooting}. When multiple fields are involved at play, four approaches were introduced in the literatures, the early trial~\cite{John:1998ip}, the undamping/damping algorithm~\cite{numericalbounce}, the wildly adopted path deformation algorithm~\cite{Wainwright:2011kj}, and the recently proposed semi-analytic perturbative approach~\cite{Akula:2016gpl}.

The generalization of~\cite{Coleman:1977py} in the case of non-zero cosmological constant allowed for both false and true vacuums was carried out in~\cite{Parke:1982pm}. The vacuum decay is enhanced in the presence of gravity when the average of two vacuums is positive and suppressed when both vacuums are negative. The vacuum decay for a positive false vacuum but negative average of two vacuums is enhanced or suppressed when the gravitational effect becomes important or not. There is currently no satisfactory description of thermal decay in curved spacetime; however, we will clarify in the future work that the gravity correction only scales as $\kappa/(TR^3)$, and it is not important because the characteristic size of bubble $R$ is much larger than the Planck length $\sqrt{\kappa}\sim1/\sqrt{G}$ for the characteristic scale of PT below the Planck scale.

\subsection{Bubble Expansion}
\label{subsec:bubble expansion}

After the bubble is nucleated within false vacuum, the bubble wall will rapidly expand until approaching the speed of light~\cite{Coleman:1977py} or colliding with other bubble walls. However, the realistic background is actually a thermal plasma full of relativistic particles \footnote{In this review, we are only interested in the GWs from PTs that happened in radiation dominated era. For GWs from PTs that happened in inflation era and matter dominated era, please refer to~\cite{Baccigalupi:1997re,Chialva:2010jt,Jiang:2015qor} and~\cite{Barenboim:2016mjm}, respectively. See also~\cite{Artymowski:2016tme} for non-standard cosmology with an additional new component that redshifts faster than radiation.}, which will interact with bubble walls by some form of frictions. The central problem of bubble expansion in plasma is to figure out the interconnections among following quantities: the bubble wall velocity $v_w$, the friction $\eta$ on the wall from plasma, the strength $\alpha$ of PT that measure the released vacuum energy density with respect to the radiation energy density, and the efficiency factors $\kappa_\phi$ and $\kappa_v$ that measure the capability of transferring the liberated vacuum energy into the bubble wall expansion and the bulk fluid motion, respectively. These quantities serve as the interface between the theoretical models of particle physics and the simulations of bubble collisions.

After the bubble is nucleated within thermal plasma, it starts stopping growing when the friction balances the net pressure from inside and outside of the bubble wall. If the pressure from outside of bubble wall is larger than the inside pressure while fluid velocity from outside of bubble wall is smaller than the inside velocity, the bubble wall behaves as deflagration, and the opposite definitions for detonation. Both deflagration and detonation are further classified as weak, Jouguet and strong types according to the fluid velocity from inside of bubble wall being smaller, equal and larger than the speed of sound. In the  paper~\cite{Steinhardt:1981ct}, it was proposed that with Jouguet condition, the velocity of bubble wall $v_w$ would be given by a simple formula~\cite{Steinhardt:1981ct} expressed in terms of the strength of PT $\alpha$ alone, so does the efficiency factor $\kappa_v$. The formula of Jouguet detonation was extensively used in the literatures at the early stage of studies of GWs from PTs despite the fact~\cite{KurkiSuonio:1995pp} that the Jouguet detonation can be unrealistic in the cosmological setup of PTs. Since the work~\cite{KurkiSuonio:1995pp}, several parallel explorations have been considered as follows.
\begin{itemize}
  \item \textit{Beyond Chapman-Jouguet condition}. Generalizing model-independent parametrization equation~\cite{KurkiSuonio:1995pp} of friction, \cite{Megevand:2009ut} was able to explore the full range of bubble wall velocity for both deflagrations and detonations, where the analytic approximations were found for both non-relativistic and ultra-relativistic wall velocities. In the state-of-art work of~\cite{Espinosa:2010hh}, it gives a unified picture and user-friendly fitting formulas for the dynamical regimes of bubble expansion. A different but more accurate approach was adopted in~\cite{Megevand:2009gh} right after~\cite{Megevand:2009ut} with microscopic considerations on the particle contents of specific models, which was revisited and improved in the subsequent work~\cite{Leitao:2012tx}. Future works can be carried out along this direction.
  \item \textit{Criteria for runaway bubble walls}. Apart from those stationary solutions of bubble wall with terminal velocity, there exists a runaway solution when the friction is too small to prevent the wall from approaching the speed of light. A simple criterion was found in~\cite{Bodeker:2009qy} that the bubble walls will runaway if the effective potential in the true vacuum remains deeper than the false vacuum even after replacing the thermal potential by its second-order Taylor expansion term in the false vacuum, but not vice versa. This was reformulated in~\cite{Espinosa:2010hh} simply by comparing the strength $\alpha$ of PTs with respect to a critical value $\alpha_\infty$. Hence, combined considerations of hydrodynamics and microphysics of specific models were later extended in~\cite{Leitao:2015ola} and~\cite{Huber:2015znp} to the runaway regime. Future works can be carried out along this direction.
  \item \textit{Reconciliations with baryogenesis}. Large GWs signals from PTs require fast-moving bubble walls to collide with each other, while baryogenesis scenarios need slow-moving bubble wall to have an effective diffusion process. In~\cite{Megevand:2009gh}, it also opens the possibility that can reconcile the baryogenesis with GWs from PTs in the presence of fermions with large Yukawa coupling and heavier stabilizing bosons. Later on,~\cite{No:2011fi} pointed out that the relevant velocity responsible for the baryogenesis is the relative velocity between the wall and the plasma, which can be much smaller than the wall velocity when the strength of PTs becomes stronger. Therefore, it is possible to simultaneously generate large GWs from PTs without jeopardizing the effectiveness of baryogenesis.
\end{itemize}

\subsection{Bubble Percolation}
\label{subsec:bubble percolation}

After bubble nucleation and bubble expansion, bubble percolation starts with colliding bubbles until PT is completed. When the initial size of nucleated bubble can be neglected, the duration of PT can be roughly estimated by the mean radius of bubbles at collisions, which is characterized by a single parameter $\beta$. Both $\beta$ and the previous mentioned $\alpha$ are evaluated at nucleation temperature $T_*$, which is defined as the temperature at which the number of generated bubble per unit time per Hubble volume is of order one. The old picture of bubble percolation consists of two sources, namely, the colliding bubble walls and the turbulent motions of bulk fluids along with their associated magnetic field. However, the new picture of bubble percolation is added with an extra source from the sound waves of bulk motion as the result of bubble collision, which exist long after the bubble percolation.
\begin{figure}
  \centering
  \includegraphics[width=0.9\textwidth]{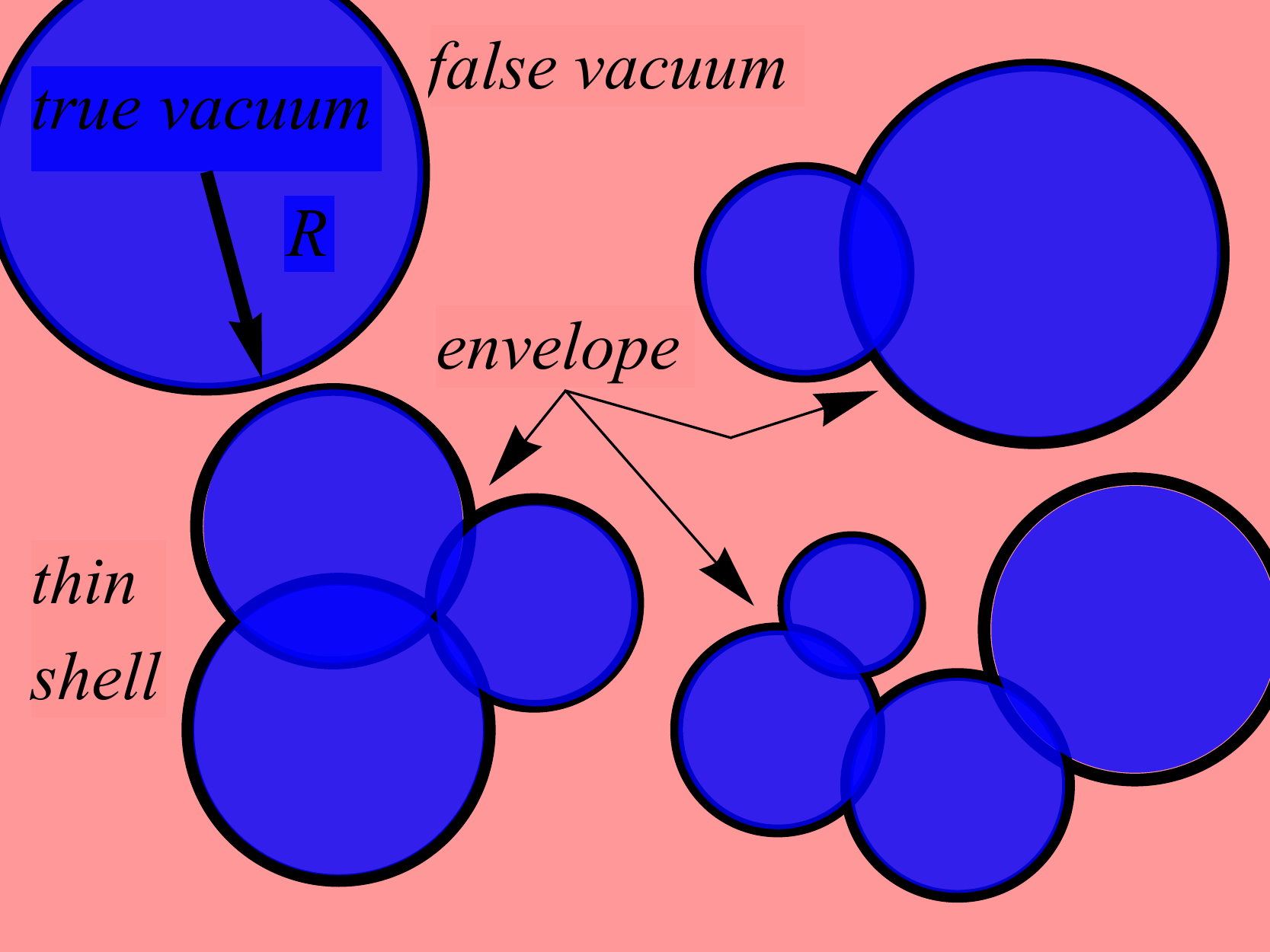}
  \caption{The schematic illustration of envelope approximation that the GWs are mainly generated from the uncollided envelopes of colliding bubble walls and any GWs from the overlap region can be neglected.}
  \label{fig:bubble}
\end{figure}

\begin{itemize}
  \item \textit{Bubble collisions}. It was first realized by Witten in~\cite{Witten:1984rs} that the QCD PTs might leave us with detectable GWs from violent bubble collisions with their peak frequency characterized by the size of bubbles when they collided and with their peak amplitude estimated by the relative size of bubbles with respect to Hubble horizon at collision. Later on, this insight was generalized by Hogan in~\cite{Hogan:1986qda} to the case of EW PTs. In a series of  papers~\cite{Kosowsky:1991ua,Kosowsky:1992rz,Kosowsky:1992vn,Kamionkowski:1993fg}, the preliminary simulations were first implemented for bubble collisions to capture the general features of GWs from PTs. It is found in~\cite{Kosowsky:1991ua,Kosowsky:1992rz} that remarkably the spectrum of GWs from simulating the collision of two vacuum bubbles in Minkowski space only depends upon the grossest features of the bubble collisions, namely, the strength $\alpha$ and duration $\beta$ of PTs, which is similar to the result of simulation for hundreds of vacuum bubbles~\cite{Kosowsky:1992vn}. As we have shown in Fig.~\ref{fig:bubble}, the envelope approximation was proposed in~\cite{Kosowsky:1991ua,Kosowsky:1992rz,Kosowsky:1992vn} that the GWs are mainly generated from the uncollided envelopes of colliding bubble walls and any GWs from the overlap region can be neglected. The extension to the thermal bubble collisions was carried out later in the simulation~\cite{Kamionkowski:1993fg}, where the Jouguet detonation mode was explicitly used and the turbulent motion in fluid stirred by bubble collisions was appreciated in this case. The state-of-the-art results of the GWs spectrum from PTs due to colliding bubble walls were settled down in~\cite{Huber:2008hg}, which although disagreed with earlier study~\cite{Caprini:2007xq}, they finally achieved consensus in~\cite{Caprini:2009fx}. Recently, under the thin-wall and envelope approximations in flat background, it was claimed in~\cite{Jinno:2016vai} that the GWs spectrum can be estimated analytically without need of  simulations. Future works can be carried out along this direction from both theoretical and numerical points of view.
  \item \textit{Turbulent MHD}. The possibility of generating GWs from turbulent motion of bulk fluid was mentioned long time ago in~\cite{Witten:1984rs} as remnant of bubble collisions, which was first estimated in~\cite{Kamionkowski:1993fg} as Kolmogorov spectrum under quadrupole approximation. Apart from above turbulent velocity field, the fully ionized plasma could also generate the turbulent magnetic field under turbulent motion, which itself is also a source of GWs. The early analysis~\cite{Kosowsky:2001xp,Dolgov:2002ra} involving MHD exhibits three problems~\cite{Binetruy:2012ze}: large-scale problem (addressed in~\cite{Caprini:2006jb,Gogoberidze:2007an}), time-evolution problem (persisted in~\cite{Gogoberidze:2007an} and addressed in~\cite{Caprini:2006jb}) and dispersion-relation problem (corrected in~\cite{Caprini:2006jb,Gogoberidze:2007an}). The state-of-art result of the GWs spectrum from PTs due to non-helical MHD turbulence was analytically established in~\cite{Caprini:2009yp}, where it ignored the circularly polarized GWs~\cite{Kahniashvili:2005qi,Kahniashvili:2008er,Kahniashvili:2008pf,Kahniashvili:2008pe,Caprini:2009pr,Kahniashvili:2009qi,Kisslinger:2015hua} from helical turbulence due to the macroscopic parity violation. The numerical simulations of relativistic MHD turbulence are certainly needed to make further confirmations for the above analytic results in the future.
  \item \textit{Sound waves}. The possibility of generating GWs from sound waves was originally pointed out in~\cite{Hogan:1986qda}, although, it left forgotten for a long time until recent revelation in~\cite{Hindmarsh:2013xza}, where envelope approximation for modeling the colliding bubble walls is abandoned and the GWs should be dominated by the overlapping sound waves in the bulk fluid. The breakthrough findings of~\cite{Hindmarsh:2013xza} were further quantitatively understood in the updated simulations~\cite{Giblin:2014qia,Hindmarsh:2015qta} and theoretically modeled in~\cite{Hindmarsh:2016lnk}, before which the possibility of having an inverse acoustic cascade was investigated in~\cite{Kalaydzhyan:2014wca}, suggesting the potentially very strong enhancement of the sound wave density at small wave number. Future works can be carried out along this direction for larger simulations with more bubbles.
\end{itemize}

\subsection{Gravitational Waves}
\label{subsec:GWPT}

The GWs from the strong first-order PTs are guaranteed through above three processes: bubble nucleation, bubble expansion and bubble percolation. The bubble nucleation requires a potential barrier in order to tunnel from the false vacuum to the true vacuum. The bubble expansion requires fast enough moving bubble walls in order to have strong signals. The bubble percolation requires efficient collisions in order to dissipate released vacuum energy into bulk fluid motions. The fitting formulas of GWs spectrums from numerical simulations are well summarized in~\cite{Caprini:2015zlo} and can be straightforwardly applied to the particle physics models, where the strength $\alpha$ and duration $\beta$ evaluated at nucleation temperature $T_*$ along with the bubble wall velocity $v_w$ are obtained from the microscopic particle physics models, whereas the efficiency factors $\kappa_\phi$ and $\kappa_v$ are approximated from the fitting formulas in~\cite{Espinosa:2010hh}. There leaves one last discussion of which particle physics models can exhibit first-order PT so that GWs can be generated.

The SM with phenomenological Higgs mechanism crosses over from the high-temperature phase to low-temperature phase~\cite{Kajantie:1996mn} if the Higgs boson mass is larger than the W boson mass. To have GWs from first-order PTs, one has to go beyond SM (BSM) \footnote{Recently, it was argued in~\cite{Ghiglieri:2015nfa} that a thermally produced GWs signal could be generated in SM physics due to shear viscosity in the plasma.}. Here, we give an incomplete list of these models with explicit discussions on GWs, which will be revisited in the near future if we want to construct the reliable templates in order to extract the GW signals from the stochastic background.

\begin{figure}
  \centering
  \includegraphics[width=0.9\textwidth]{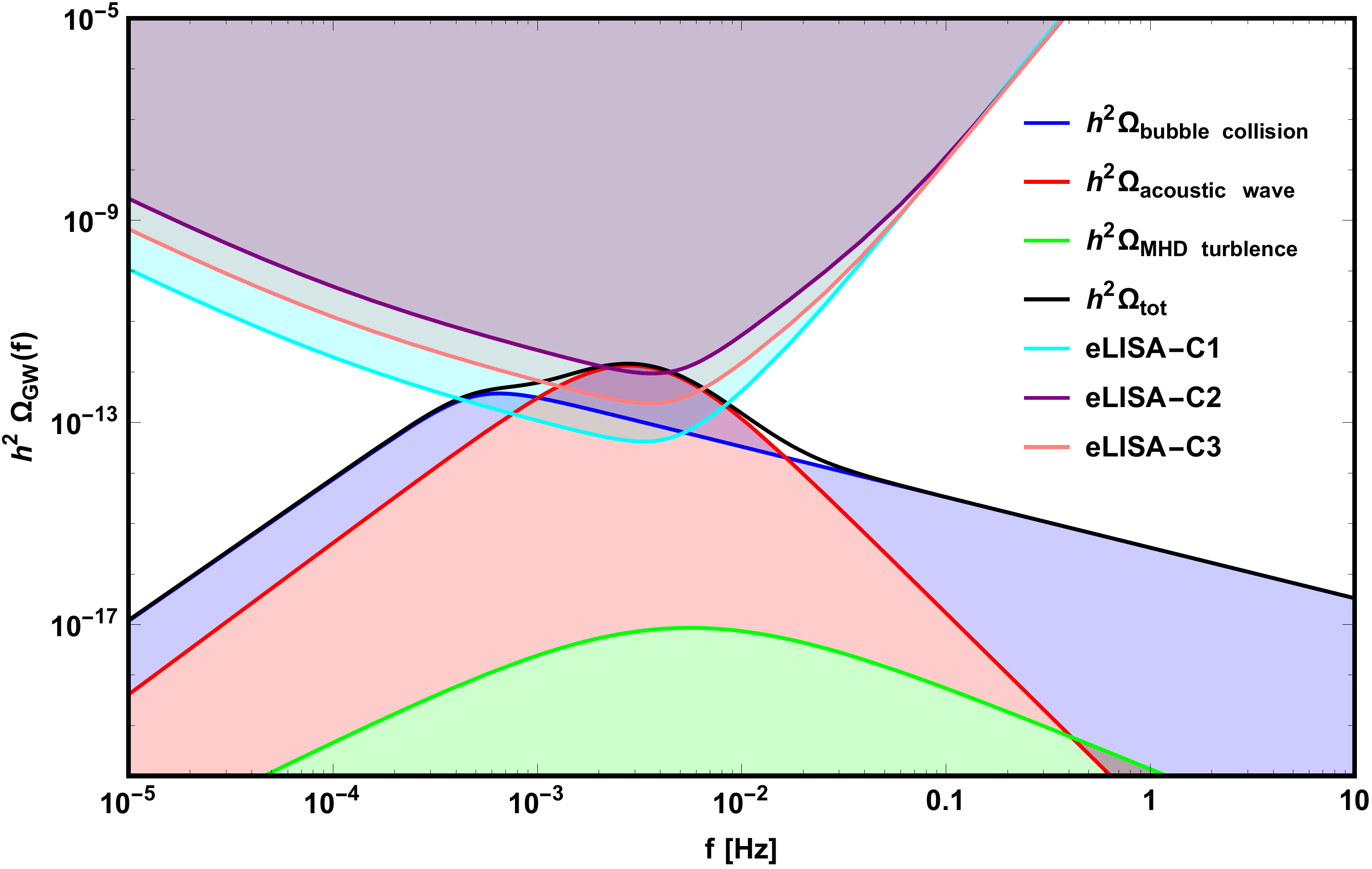}
  \caption{The spectrum of GWs from PTs that originated from dimension-six operator with respect to the sensitivity curves of various configurations of eLISA project~\cite{Binetruy:2012ze}.}
  \label{fig:GWPT}
\end{figure}

\begin{itemize}
  \item \textit{Higher dimensional operators}. The simplest example is to add a cubic term, which is usually expected in the supersymmetric extensions of SM in order to make a strong first-order PTs. By adopting the polynomial fitting formulae~\cite{Adams:1993zs} for the bounce action from a general quartic potential with a cubic term, a semi-analytic calculation of the GW signal from the EWPT was carried out in~\cite{Kehayias:2009tn}. The other important example is the dimension-six operator~\cite{Grojean:2004xa,Bodeker:2004ws}, of which the GW signals from PTs were first analyzed in~\cite{Huber:2007vva} and revisited in~\cite{Delaunay:2007wb} with full scope of one-loop effective potential at finite temperature. In both examples, the bubble walls with detonations or run-away behavior have been considered in~\cite{Leitao:2015fmj}. We produce in  Fig.~\ref{fig:GWPT} the spectrum of GWs from PTs that originated from the dimension-six operator, which will be discussed in details in a future work~\cite{wang:future}.
  \item \textit{Additional scalar sectors}. The simplest and extensively studied example is the gauge singlet scalar extension of SM~\cite{Leitao:2012tx,Jinno:2015doa,Hashino:2016rvx,Huang:2016cjm,Hashino:2016xoj,Balazs:2016tbi,Vaskonen:2016yiu,Beniwal:2017eik}, which can be naturally fitted into previously mentioned cubic term~\cite{Kehayias:2009tn} and can be tested at future colliders by measuring the triple Higgs coupling precisely \footnote{It has been exemplified in~\cite{Huang:2016odd} with dimension-six operator that the future space-based interferometers could probe the parameter spaces that are unreachable for the current particle colliders, but can have overlap with the future particle colliders, therefore a cross-check is possible.}. The other important example is the charged scalar under SM gauge group, of which the simplest realization is the two-Higgs-doublet model (2HDM). The GW from 2HDM was preliminarily analyzed in~\cite{Kakizaki:2015wua} and further studied in~\cite{Caprini:2015zlo} for the case of CP-conservation and recently revisited in~\cite{Dorsch:2016nrg} for the case of CP-violation.
  \item \textit{Supersymmetric extensions}. The capability of detecting GWs from PTs in Minimal Supersymmetric Standard Model (MSSM) and Next-to-Minimal Supersymmetric Standard Model (NMSSM) were first estimated in~\cite{Apreda:2001tj} and further explored in~\cite{Apreda:2001us}. Both the PTs from MSSM~\cite{Leitao:2012tx} and NMSSM~\cite{Huber:2007vva} were thought to be not strong enough to produce significant GW signals; however, the parameter space with strong first-order PTs has been identified for a modified version of MSSM~\cite{Garcia-Pepin:2016hvs} and a general version of NMSSM~\cite{Huber:2015znp}.
  \item \textit{Hidden dark sectors}. The cosmological implications concerning with possible production of GWs from hidden dark sector were first explored in~\cite{Espinosa:2008kw}, and later discussed for light GeV scalar~\cite{Das:2009ue}, vector thermal dark matter~\cite{Hambye:2013sna}, UV-conformal dark sector~\cite{Dorsch:2014qpa}, $SU(N)$ dark sectors with $n_f$ flavors~\cite{Schwaller:2015tja}, dark $U(1)$ gauge complex scalar singlet~\cite{Jaeckel:2016jlh}, hidden sector with run-away bubble walls~\cite{Katz:2016adq}, and PTs involving successive hidden gauge symmetry breaking~\cite{Huang:2017laj}, dark matter asymmetry~\cite{Baldes:2017rcu} and two-step transition~\cite{Chao:2017vrq}. The GWs from first-order PTs could be a unique probe for these dark hidden sectors.
  \item \textit{Other BSM extensions}. The GWs from first order PTs were also analyzed in the case of extra dimensions~\cite{Randall:2006py,Nardini:2007me,Konstandin:2010cd,Konstandin:2011dr},  Peccei-Quinn PT~\cite{Dev:2016feu}, non-linear EWPT~\cite{Kobakhidze:2016mch} and QCD PT~\cite{Boeckel:2009ej,Schettler:2010dp,Boeckel:2011yj,Anand:2017kar}.
\end{itemize}

\section{The Gravitational Waves From Binary Systems}
\label{sec:binary system}

GW sources can be divided into two categories, one is deterministic and the other is stochastic. For example, the primordial GW produced by the quantum fluctuations and by PTs during the early Universe belongs to the stochastic one. Due to the intrinsic random character, we cannot predict the waveform produced by the stochastic sources. The deterministic sources include two types. One type is predictable, while the other one is not. For example, the supernova is believed producing GWs. But the dynamics of supernova is too complicated to predict the related GW form. Another example is the foreground noise of LISA produced by white dwarf-white dwarf binaries in our milky way galaxy~\cite{Stroeer:2006rx,Ruiter:2007xx,Crowder:2006eu}. Due to the large number of such binaries, the combination of these GW signals makes the waveform unpredictable. In this section, we focus on the predictable GW sources. For such sources, we can construct theoretical model for gravitational waveform before GW detection experiment~\cite{cao2016numerical}. Then when the experiment data are ready, we can use matched filtering data analysis techniques to improve the GW detection sensitivity. And more, the theoretical model can be used to realize the astronomy detection of the GW sources through matched filtering~\cite{cao2016gravitational,cai2016gravitational}.

\begin{figure}
  \centering
  \includegraphics[width=0.9\textwidth]{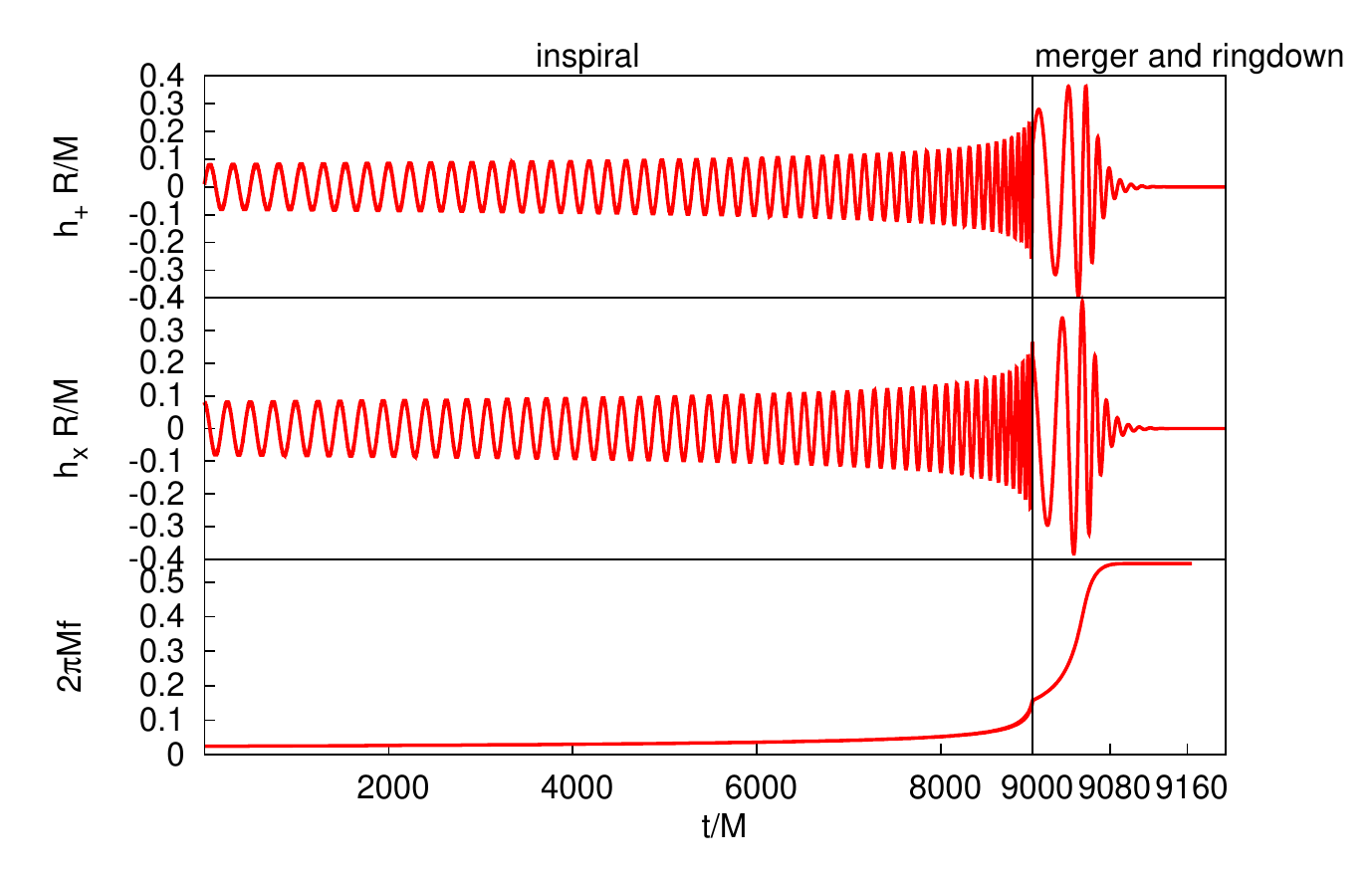}
  \caption{Typical example of the gravitational waveform and the corresponding frequency. In this example, the binary system is a binary black hole with equal mass. And the two black holes are spinless. The whole evolution process of a binary black hole system includes inspiral, merger and ringdown. But the boundary to distinguish these three stages is fuzzy. Compared to the inspiral stage, the merger and ringdown stages are much shorter. In this plot, we use different time scale for inspiral and merger/ringdown stage in order to make the merger/ringdown stage clearer.}\label{fig:GWform}
\end{figure}

Binary systems are among the most important and major sources for GW detection projects including PTA, space-based laser interferometer detectors such as eLISA, Taiji and Tianqin projects, ground-based laser interferometer detectors such as Advanced LIGO, Advanced Virgo, Kagra and others. The GW frequency of binary system around merger state can be characterized by the ISCO (most Inner Stable Circular Orbit) frequency:
\begin{align}
f_{isco}=\frac{1}{6^{3/2}2\pi M},
\end{align}
where $M$ is the total mass of the binary system, and we have used geometric unit with $c=G=1$. When $M=M_\odot$, $f_{isco}\approx2000$Hz. At the same time, binary systems belong to predictable GW sources~\cite{cai2016gravitational}. A typical example of the gravitational waveform and the related frequency are shown in Fig.~\ref{fig:GWform}. Note the time scale in the plot is different for inspiral stage and the merger/ringdown stage. The component of the binary system can be white dwarf, neutron star, black hole and even quark star. If gravitational theory beyond general relativity is true, other objects including axion star~\cite{Raby:2016deh}, gravastar~\cite{Visser:2003ge} and other exotics might also be the component of binary system.

Based on current understanding of stellar evolution, the final fate of a star may be white dwarf, neutron star or black hole which is determined by the mass of the star. Binary systems can be formed through capture process or many body interaction. Due to the gravitational slingshot effect involved in three-body interaction, binary systems can be formed through many body interaction. Because a binary system is dynamically stable,  binary systems are expected to be quite common in our Universe.

The direct detection of GWs opens the GW astronomy. Through this new window to our Universe, GWs will not only bring us many new observations on kinds of objects in the Universe, but also give us many insights on fundamental physical theory. For example, black hole is a mystery from quantum gravity theory view. Especially, black hole presents us an information loss puzzle. Is it possible the black hole horizon be replaced with other objects such as firewall or anything else? Or any experiments present evidence that black hole horizon really exists? Unfortunately, one cannot get observational proof of the existence of a black hole event horizon based on EM waves~\cite{Abramowicz:2002vt}. In contrast, it is possible to use GW observation to give a direct proof of the existence of a black hole event horizon~\cite{Abedi:2016hgu}. Currently, these promising aims can be most possibly achieved by predictable GW sources, especially binary systems. In order to realize these aims, we have to construct accurate enough gravitational waveform model for binary systems. There are three kinds of methods available to treat binary systems. They are PN approximation, numerical relativity and black hole perturbation method which are widely used for early inspiral stage, plunge and merger stage, and post merger stage, respectively.

\subsection{Post-Newtonian Approximation}
\label{subsec:PN}

Based on quadrupole formula~\cite{baumgarte2010numerical}, we can estimate the GWs related to a binary system:
\begin{align}
h_+&=-\frac{M}{r}2\Omega^2R^2(1+\cos^2\theta)\cos[2\Omega(t-r)],\\
h_\times&=-\frac{M}{r}4\Omega^2R^2\cos\theta\sin[2\Omega(t-r)],
\end{align}
where $(r,\theta,\phi)$ is the position of the observer with respect to the binary, $M$, $R$ and $\Omega$ are the total mass, the separation and the mutual orbiting frequency of the two components of the binary. Here $h_+$ and $h_\times$ correspond to  two polarization modes of GWs. These two modes are related to the dynamical metric form as
\begin{align}
\mathrm{d}s^2=-\mathrm{d}t^2+\mathrm{d}r^2+r^2(1+h_+)\mathrm{d}\theta^2+r^2\sin^2\theta(1-h_+)\mathrm{d}\phi^2+2r^2\sin\theta h_\times \mathrm{d}\theta \mathrm{d}\phi,
\end{align}
with $(t,r,\theta,\phi)$ corresponding to the transverse-traceless gauge~\cite{liang00}. Here we need point out two points. First one is that the planer wave approximation has been taken in the above equation which is reasonable because the field point is very far away from the source. The second is that above estimation about $h_+$ and $h_\times$ is only leading order of the PN approximation which is reasonable because the weak GW source moves slowly. For most realistic GW sources, these approximations will break down. Even for cases in which the approximation is reasonable, above approximation cannot satisfy the need of GW detection~\cite{Lindblom:2008cm,Lindblom:2016csg}. For example, when the two components of a binary system are some far away, they move slowly. So, this kind of situation corresponds to weak and moving slowly GW source. Although the above approximation is reasonable, the accuracy is not good enough. Then, more PN order corrections are needed to improve the accuracy~\cite{Blanchet:2013haa}.

The theoretical framework of the PN approximation for the processing of binary systems consists of two parts: the dynamics of binary components and the GWform. In order to construct the GW model, we need to solve the binary dynamics, and then put the solution into the waveform theory part to get the explicit GW model. In general, the dynamical equations of the PN approximation are a highly nonlinear system of ordinary differential equations. In that case, the numerical method has to be employed to get the solution. GW models in which the GWform is expressed in time domain include TaylorT2, TaylorT4 and others~\cite{Buonanno:2009zt}. Through stationary phase approximation, the post-Newtonian equations can be transformed into a frequency domain problem, and fortunately an approximate analytical expression can be obtained. Such a model is typically represented by the TaylorF2 model. Corrected by the binary black hole spin dynamics, especially the effect of the orbital precession, the TaylorF2 model is modified and improved by single and double precession model~\cite{Lundgren:2013jla,Chatziioannou:2014bma}. The TaylorT2 model is replaced by the X model when  the elliptic orbit of the binary system is considered~\cite{Hinder:2008kv}. The theoretical model of the TaylorF2 model is extended to the elliptic orbit binary system including the post-circular model~\cite{Yunes:2009yz} and the enhanced post-circular model~\cite{Huerta:2014eca,Sun:2015bva}.

\subsection{Numerical Relativity}
\label{subsec:numrical}

\begin{figure}
  \centering
  \includegraphics[width=0.9\textwidth]{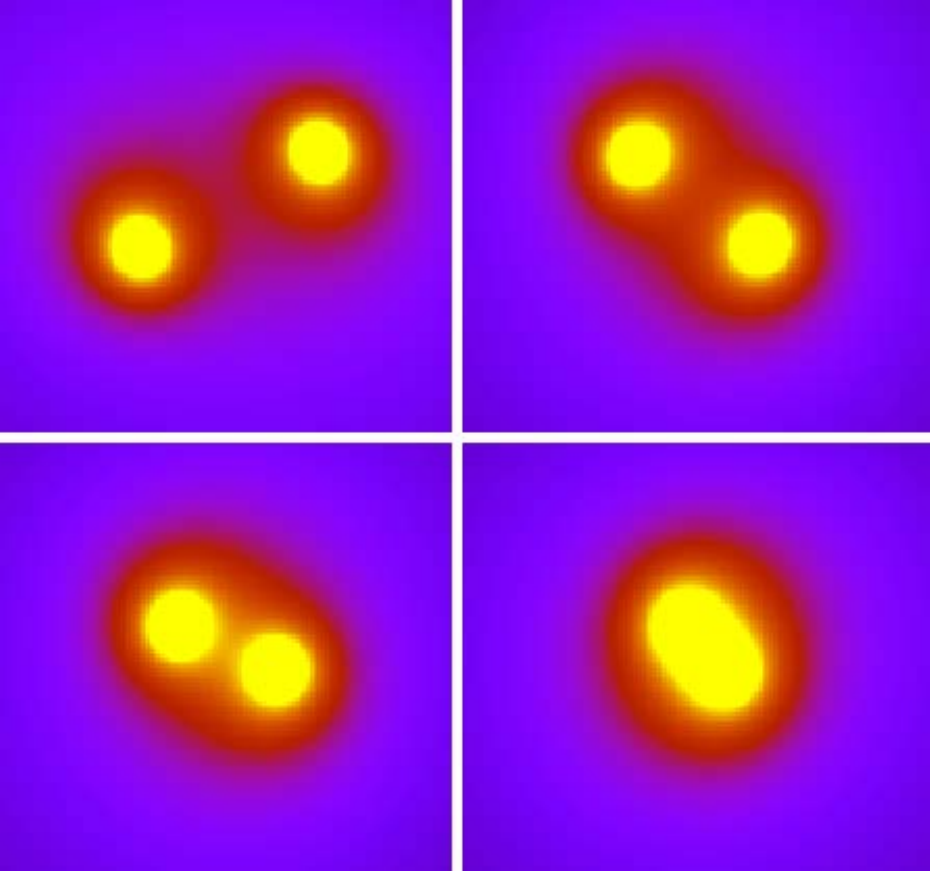}
  \caption{Typical evolution of binary black hole simulated by numerical relativity. The yellow region corresponds to the highly curved region which is roughly the position of black holes. The four snapshots correspond to inspiral, merger and ringdown stages. This result is simulated by AMSS-NCKU software.}\label{fig:BBH}
\end{figure}

Applying numerical methods to solve the Einstein equation is the topic of numerical relativity. Currently, the numerical relativity can only  deal with the problem of GW modeling for the plunge and merger stage of binary systems. Because numerical relativity solves the Einstein equation without any approximation up to numerical error, this feature makes numerical relativity a universal tool to address kinds of GW sources. In Fig.~\ref{fig:BBH}, we show an example of the evolution process of a binary black hole system simulated by numerical relativity. But the nature of diffeomorphism invariance of the general relativity brings numerical calculation special difficulties. The study of numerical relativity began in the 60s of the last century~\cite{HAHN1964304}, but the numerical instability made the code collapse after a few calculation steps. In 1990s, the construction of LIGO hardware strongly demanded the GW source modeling. So tens of universities and research institutions in USA and Europe jointly launched a Binary black hole Grand Challenge Project. However, the stability problem of numerical relativity has not been solved by that project. Out of one's expectation, the stability problem of numerical relativity was solved for the first time by Pretorius in 2005~\cite{Pretorius:2005gq}. In the following 2006, Baker group and Campanelli group also independently solved the instability problem~\cite{Campanelli:2005dd,Baker:2005vv}. Till now, there are more than ten numerical relativity groups around the world that have solved the stability problem, including the Princeton University, California Institute of Technology, University of Jena, Max Planck institute, the academy of mathematics and systems science, CAS~\cite{Cao:2008wn} and others.

It is worth pointing out that within kinds of stable numerical relativity methods, which tips and treatments are necessary and/or sufficient for stable computation are still an open question. From the  viewpoint of the theory of partial differential equations, one can only apply the hyperbolicity analysis to linearized Einstein equation~\cite{Hilditch:2013sba}. In practice, however, the success of Pretorius in 2005 depends much on the formalism of the Einstein equations and the adaptive mesh refinement technique. Regarding the formalism, the so-called BSSN equation and the generalized harmonic coordinate equation have been widely used in numerical relativity. Adaptive mesh refinement is a very effective method to deal with multiscale problems. At present, the parallel adaptive mesh refinement codes developed especially for Einstein equations include the BAM code (developed by Bruegmann), AMSS-NCKU code (developed by Zhoujian Cao)~\cite{AMSSNCKU}, PAMR code (developed by Pretorius) and the Carpet code (developed by Schnetter).

Besides the stability problem, the major issues involved in numerical relativity are accuracy and efficiency. As for the formalism factor, more and more investigations show that Z4c and CCZ4 formalism is better than the BSSN formalism~\cite{Bernuzzi:2009ex,Cao:2011fu,Alic:2013xsa}. As an example of partial differential equation formalism of Einstein equations, we show Z4c formalism as following, which is proposed first time in 3D form in \cite{Cao:2011fu}:
\begin{align}
\partial_t\chi&=\frac{2}{3}\chi[\alpha(\hat{K}+2\Theta)-D_i\beta^{i}],\\
\partial_t\tilde\gamma_{ij}&=
- 2 \alpha \tilde A_{ij}+2\tilde\gamma_{k(i}\partial_{j)}\beta^k
- \frac{2}{3} \tilde\gamma_{ij}\partial_{k}\beta^k
+ \beta^k\partial_k\tilde{\gamma}_{ij},\\
\partial_t\hat{K}&=-D_iD^i\alpha + \alpha[\tilde{A}_{ij}\tilde{A}^{ij}
+\frac{1}{3}(\hat{K}+2\Theta)^2\nonumber\\
&+\kappa_1(1-\kappa_2)\Theta]
+4\pi\alpha[S+\rho_{\textrm{\tiny{ADM}}}]+\beta^i\partial_i\hat{K},\\
\partial_t\tilde{A}_{ij}& = \chi [-D_iD_j\alpha+\alpha(R_{ij}
-8\pi S_{ij})]^{\textrm{tf}}
+\alpha [(\hat{K}+2\Theta)\tilde A_{ij} \nonumber\\
&- 2\tilde A_{ik}\tilde A^k{}_j]+2\tilde A_{k(i}\partial_{j)}\beta^k
-\frac{2}{3}\tilde A_{ij}\partial_k\beta^k
+ \beta^k\partial_k\tilde A_{ij},\\
\partial_t\Theta&=\frac{1}{2}\alpha\big[R-\tilde{A}_{ij}\tilde{A}^{ij}
+\frac{2}{3}(\hat{K}+2\Theta)^2-16\pi\rho_{\textrm{\tiny{ADM}}}\nonumber\\
&-2\kappa_1(2+\kappa_2)\Theta\big]+\beta^i\partial_i\Theta,\\
\partial_t\tilde{\Gamma}^i&= \tilde{\gamma}^{jk}\partial_j\partial_k\beta^i
+\frac{1}{3}\tilde{\gamma}^{ij}\partial_j\partial_k\beta^k
-2\tilde{A}^{ij}\partial_j\alpha\nonumber\\
&+2\alpha \big[\tilde{\Gamma}^i{}_{jk}\tilde A^{jk}
-\frac{3}{2}\tilde{A}^{ij}\partial_j\ln\chi
-\frac{1}{3}\tilde{\gamma}^{ij}\partial_j(2\hat{K}+\Theta)\nonumber\\
&-\kappa_1(\tilde{\Gamma}^i-\tilde{\Gamma}_{\textrm{\tiny{d}}}{}^i)
-8\pi\tilde{\gamma}^{ij}S_j
\big]+\frac{2}{3}\tilde{\Gamma}_{\textrm{\tiny{d}}}{}^i\partial_j\beta^j
-\tilde{\Gamma}_{\textrm{\tiny{d}}}{}^j\partial_j\beta^i\nonumber\\
&+\beta^j\partial_j\tilde{\Gamma}^i.
\end{align}
The unknown functions which need to be solved numerically are listed in the left-hand side of the above equations. For the meaning of the notations used in the above equations, we refer our readers to \cite{Cao:2011fu}. Spectral method, finite difference method and finite element method are three categories of numerical methods for partial differential equations. The vast majority of numerical relativity groups take the finite differential method. The SpEC code of California Institute of Technology uses spectral method. On the other hand, the application of finite element method into  numerical relativity is still very few~\cite{Cao:2015via}. The spectral method has good computational efficiency due to its exponential convergence. However, the characteristics of its global data exchange limits its ability of parallel computing. The finite difference method working with domain decomposition algorithm can achieve good parallel computing efficiency. In order to deal with the multiscale characteristics of astrophysics, the refinement of the multilayer data structure is essential in the finite difference method. However, this method limits the number of cores for parallel computation by a single data layer, thus limits the strong parallel scalability. In contrast, the finite element method can combine the exponential convergence of the spectral method and the high parallel scalability of the difference method. Therefore, in principle, it is expected that the finite element calculation of the Einstein equations can achieve good strong parallel scalability. However, it is still an issue to construct the weak form of the Einstein equations, and to realize large-scale scientific computation~\cite{cao2016numerical,cai2016gravitational}.

Now the major challenges to numerical relativity placed by GW source modeling are the calculation of binary systems with mass ratio more than 100 to 1 and the well-converged simulation of binary systems including neutron stars~\cite{cao2016gravitational}.

\subsection{Black Hole Perturbation}
\label{subsec:BHperturb}

After the merger of the binary components, the system will ringdown and finally settle down to a Kerr black hole. If the two components are white dwarf and/or neutron star, the final remnant could  be a stable neutron star. Here, we only concern with the case with Kerr black hole as the final product. Then the ringdown stage can be described by perturbation of a Kerr black hole. The black hole perturbation theory was pioneered by Regge, Wheeler and Zerilli~\cite{Regge:1957td,Zerilli:1971wd} in which perturbation around a Schwarzschild black hole  was considered. Teukolsky investigated the perturbation of a Kerr black hole on spacetime curvature level~\cite{Teukolsky:1973ha,Teukolsky:1974yv}. The former is called the Regge-Wheeler-Zerilli equation, and the latter is called the Teukolsky equation. The perturbation method used by Teukolsky is also valid to Schwarzschild black hole. For Schwarzschild black hole, Regge-Wheeler-Zerilli method and Teukolsky method is equivalent~\cite{chandrasekhar1998mathematical}. In the early 70s of the last century, some scholars have begun to use the Regge-Wheeler-Zerilli equation to study the associated GW for a test particle falling into a black hole~\cite{Davis:1972ud,Ruffini:1973ky,Fujita:2004rb}. Because the Teukolsky equation admits spin of a black hole, it is valid more extensively than the Regge-Wheeler-Zerilli equation. In recent years, there have been many works using the Teukolsky equation to calculate the GWs.

The Teukolsky equation can be solved by the PN approximation method (note that the Teukolsky equation is the target in question here, instead of Einstein equation itself like what described in the above subsection) or through numerical method. The PN approximation method was mainly developed by Mano, Suzuki, Takasugi and others. These authors developed some hypergeometric functions and Coulomb wave functions to approximate the homogeneous solution of the Teukolsky equation~\cite{Tagoshi:1996gh,Sasaki:2003xr}. Later on, Fujita and his coworkers applied PN approximation method to these analytical solutions to calculate the GWform and the related energy flux for extreme mass ratio binary systems. Currently, 22PN accuracy for Schwarzschild black hole and 11PN accuracy for Kerr black hole have been achieved~\cite{Fujita:2012cm,Fujita:2014eta}. However, such accuracy  still do not yet satisfy the requirement of GW detection experiment~\cite{Sago:2016xsp}.

The numerical methods to solve the Teukolsky equation can be divided into two categories: frequency domain method and time domain method. The frequency domain method divides the original Teukolsky equation into two ordinary differential equations through variable separating method and the Fourier expansion~\cite{Teukolsky:1973ha}. Regarding the radial equation, we can transform it to Sasaki-Nakamura equation before solving it numerically~\cite{Sasaki:1981sx}. The Teukolsky equation can only provide the energy flux and the GWs. In order to provide the orbit of the test particle, Hughes and his coworkers applied adiabatic approximation method to the geodesic equation to treat the inspiral~\cite{Hughes:1999bq}. Now this kind of approximation has become an important method to treat binary systems with extreme mass ratio. Different from the usual finite difference numerical method, Fujita and Tagoshi used the hypergeometric function and the Coulomb wave function to expand the original equations which gives a quicker and more accurate numerical method~\cite{Han:2016zee}. The numerical method in time domain needs to solve a set of 2+1 partial differential equations. At present, one mainly uses the finite difference method to solve the problem (but see~\cite{Sopuerta:2005gz} for the finite element method). Due to the extreme mass ratio of the involved binary system, the related GW signal may last several years. Although the computation required by Teukolsky equation is much simpler than numerical relativity, how to improve the computational efficiency of the Teukolsky equation is also a great challenge for GW source modeling.

\subsection{Cosmological Probes}
\label{subsec:cosmoprobe}

In 1986, Schutz showed that from the GWs one can infer the Hubble constant by using the fact that the GWs from the binary systems encode the absolute distance information~\cite{Schutz:1986gp}. The inspiraling and merging compact binaries consisting of neutron stars and black holes can be considered as standard candles, or ``\textit{standard sirens}''. The name of ``siren'' is due to the fact that the GW detectors are omni-directional and detect coherently the phase of the wave, which  makes them in many ways more like microphones for sound than like conventional telescopes. For an expanding Universe, the chirp waveform of the GW can be generalized to the cosmological case by multiplying all masses by the factor $1+z$ and the physical distance $D$ can be replaced by the luminosity distance $d_L$~\cite{Krolak:1987ofj,Sathyaprakash:2009xs}.
In fact, the waveform $h$ produced by compact binary inspirals during the inspiral phase is theoretically well described by the analytical solution:
\begin{align}
	h_\times = \frac{4}{d_L(z)} \left( \frac{G \mathcal{M}_c(z)}{c^2} \right)^{5/3} \left( \frac{\pi f}{c} \right)^{2/3} \cos\iota \sin[\Phi(f)] \,,
	\label{eq:gwf}
\end{align}
which is valid at the lowest (Newtonian) order for the ``cross'' GW polarization (the same expression holds for the ``plus'' polarization with a difference dependence on the orientation of the binary's orbital plane $\iota$). Here, $\mathcal{M}_c(z)$ is the (redshifted) chirp mass, $f$ the GW frequency at the observer and $\Phi(f)$ its phase. For the standard sirens, the luminosity distance $d_L(z)$ is the most important parameter entering the waveform \eqref{eq:gwf}. Parameter estimation over the observed GW signal \eqref{eq:gwf} can directly yield the value of the luminosity distance of the GW source, together with an uncertainty due to the detector noise. Once the signal is detected and characterized, the luminosity distance of the source can be extracted. This indicates that one can measure the luminosity distance directly without the need of the cosmic distance ladder: standard sirens are ``\textit{self-calibrating}''. However, it is not possible to infer the redshift $z$ from the GW signal of the binary of black holes because all of the observed parameters such as the masses and distance are redshifted by the same factor $1+z$. To use the distance measurement for cosmography, one has to obtain the redshifts of the GW sources by some independent methods.

Before considering how to obtain the redshift information, one should think about how accurately the distance can be measured. The performance of the standard siren is limited by several effects. Firstly, the intrinsic measurement uncertainty in the amplitude of the detector's response is simply the inverse of the signal-to-noise ratio (SNR)~\cite{Sathyaprakash:2009xt,Cai:2016sby}, which is related to the detector sensitivity. Second, the weak gravitational lensing will distort the measurement of the luminosity distance, producing a magnification or demagnification on the order of a few percent~\cite{Bartelmann:1999yn,Holz:2005df,Dalal:2006qt,Jonsson:2006vc}. Furthermore, the largest contribution of the uncertainty comes from the limited direction and source orientation sensitivity, and there is a large correlation between the distance, the sky position and orientation of the source. Thus, a network of detectors is needed to measure the position and orientation of the binary and to break the degeneracy among  these parameters. By observing a simultaneous detection of a beamed EM signal, one can determine the sky position accurately and also will help to improve the measurements of the distance. The EM signal, called the EM counterpart, is also an important issue to the GW on cosmography.

The most traditional way to obtain the redshift of a  GW event is through an accompanying EM signal, the EM counterpart. The binary merger of an NS with either an NS (BNS) or BH (BHNS) is hypothesized to be the progenitor of a short and intense burst of $\gamma$-rays, a so-called SGRB~\cite{Nakar:2007yr}. An EM counterpart like SGRB can provide the redshift information if the host galaxy of the event can be pinpointed. Moreover, SGRBs are likely to be strongly beamed phenomena~\cite{Nakar:2005bs}, which allow one to constrain the inclination of the compact binary system, breaking the distance-inclination degeneracy. The GWs with short $\gamma$-rays bursts or other EM counterparts as the standard sirens have been studied in various  papers (see \cite{Markovic:1993cr,Holz:2005df,Kocsis:2005vv,Dalal:2006qt,MacLeod:2007jd,Linder:2007ge,
Stavridis:2009ys,Cutler:2009qv,Arun:2008xf,Sathyaprakash:2009xt,Nissanke:2009kt,Zhao:2010sz,
DelPozzo:2011yh,Petiteau:2011we,Taylor:2012db,Arabsalmani:2013bj,Tamanini:2016zlh,Yu:2016tar,
Caprini:2016qxs,Cai:2016sby}, and references therein). For example, Nissanke \textit{et.al.}~\cite{Nissanke:2009kt} used MCMC method and found that a network of advanced LIGO detectors can constrain the Hubble constant to a $5\%$ accuracy. In Ref.~\cite{Zhao:2010sz}, the authors demonstrate that with $1000$ GW events detected by ground-based Einstein Telescope~\footnote{Website for ET: \href{www.et-gw.eu}{www.et-gw.eu}} it is possible to constrain the Hubble constant and dark matter energy parameter up to  $\Delta h_0\sim 5\times 10^{-3}$ and $\Delta \Omega_m\sim 0.02$ using Fisher matrix approach, which was also found by Cai {\it et al.}~\cite{Cai:2016sby} using MCMC method. Furthermore, Cai {\it et al.} used the Gaussian Process method and found that the equation of state of the dark energy can be constrained to $\Delta w(z) \sim 0.03$ in the low redshift region, which gives  a better constraint than Ref.~\cite{Zhao:2010sz}. For the space-based detector LISA~\footnote{Website for eLISA: \href{www.elisascience.org}{www.elisascience.org}}, the expansion of the Universe and interacting dark energy have also been studied by Refs.~\cite{Tamanini:2016zlh,Caprini:2016qxs}.

There also exist other methods to infer the redshift information of the GW events, such as galaxy catalog proposed by Schutz~\cite{Schutz:1986gp}, neutron star mass distribution~\cite{Markovic:1993cr,Taylor:2011fs}, and the tidal deformation of neutron stars~\cite{Messenger:2011gi,DelPozzo:2015bna}. Measuring the redshift associated with a GW event is one of the biggest challenges in the future, see~\cite{Namikawa:2015prh}; however, for a new GW standard sirens probe without redshift information by utilizing those BH binaries to be a tracer of the large-scale structure~\cite{Namikawa:2016edr}. In addition, the spin of BH can also help us estimate the GW's parameters. GWs as the standard sirens to probe the cosmological parameters can provide an independent and complementary alternative to current experiments. It is expected that combining GW data with other astronomical observations such as supernovae, and adopting a better data analysis approach, the cosmological parameters could be constrained more precisely than the current situation.

\section{Conclusions}
\label{sec:conclusion}

The direct detection of GWs by LIGO initiates a new era of GW astronomy and GW cosmology. The GW physics is not only related to gravitational physics, but also closely related to particle physics, cosmology and astrophysics. The GWs provide us a new powerful tool to reveal various secrets of the nature. Indeed, a lot of relevant papers have appeared since the announcement of the direct detection of GWs. In this paper, we have briefly introduced three kinds of GW sources and relevant physics. They are GWs produced during inflation and preheating in the early Universe, from cosmic PTs and dynamics of compact binary systems, respectively. We also have discussed in a simple way the GWs as standard siren in the evolution of the Universe. Due to the limitation of space, we are not able to discuss all aspects of GW physics, but only focus on some main issues. Of course, it is also impossible to list a complete list of the references, quite probably some important references are missed here, we should apologize for this.

\acknowledgments
This work is supported in part by the National Natural Science Foundation of China Grants No.11690021, No.11690022, No.11690023, No.11622546, No.11575272, No.11375260 and No.11335012, in part by the Strategic Priority Research Program of CAS Grant No.XDB23030100 and by the Key Research Program of Frontier Sciences of CAS Grant No.QYZDJ-SSWSYS006.

\bibliographystyle{JHEP}
\bibliography{ref}

\end{document}